\documentclass[doublecol]{epl2} 

\usepackage{amsmath,amsthm,latexsym,amssymb,amsfonts,epsfig,graphicx} 
\usepackage{color}

\graphicspath{{figures/}}

\newcommand{\beq}{\begin{equation}}
\newcommand{\eeq}{\end{equation}}

\usepackage[T1]{fontenc}        

\title{Diagnosing a two-body state of ultracold atoms with light}
\shorttitle{Diagnosing a two-body state of ultracold atoms with light} 
\author{Rafa{\l} O{\l}dziejewski\inst{1}  \and Kazimierz Rz\k{a}\.{z}ewski\inst{1}}
\shortauthor{Rafa{\l} O{\l}dziejewski \etal}

\institute{                    
  \inst{1} Center for Theoretical Physics, Polish Academy of Sciences, Al. Lotnik\'{o}w 32/46, 02-668 Warsaw, Poland
}

\abstract{
An absorption of a weak pulse by two identical atoms moving in a trap is investigated. Based on atom-light interactions we present a microscopic model of a two-body wave function diagnosis. We study the influence of pulse properties on the results. We show that a pulse duration impact a resulting one-photon and two-photons absorption probabilities significantly.
}

\begin{document}

\maketitle

\section{Introduction}
The enormous development in the field of ultracold atoms in the last 20 years is not only due to the advanced cooling methods \cite{Chu1998,Phillips1998,Cohen1998}, but also to the imaging techniques. A diagnosis of a Bose-Einstein condensate is always based on atom-light interactions. The multitude of possibilities created by this phenomenon allows researchers to deeply examine various aspects of quantum systems at extremely low temperatures \cite{Anderson1995,Davis1995, Ketterle2002, Cornell2002}.\\
\indent One of the most widespread experimental method of investigation of ultracold gases is the absorption imaging. The key role in this technique plays a fact that the absorption rate is proportional to the column density of atoms, or so is tacitly assumed. Sending a resonant probe light pulse through a sample and observing the "shadow" left by the atomic absorption with a CCD camera opens access to many relevant properties like density profiles of atomic clouds and higher order correlation functions \cite{grondalski1999,altman2004,folling2005,greiner2005,esteve2006}. Recently it was even possible to image a single atom \cite{streed2012}. Typically a weak pulse is used to avoid multiple scatterings and therefore enhance control over the measurement.\\
\indent Although the general principles of absorption imaging are well understood, there remain some open questions waiting to be addressed. Recently a process of splitting of the Bose-Einstein condensate was analysed \cite{gorski2015}. Within the classical field approximation, see for instance \cite{witkowska2009,brewczyk2013}, it was shown that the statistical properties of the condensate depend on the observation time. Whereas it is possible that these findings are not exactly related to a real quantum measurement with light, a role of spatial and temporal properties of a light pulse in the absorption imaging, to the best of our knowledge, was not analysed in detail. Well controlled experiments with just a few trapped atoms are now possible \cite{Serwane2011,Baier2016}. Such a simple system offers the unique opportunity to scrutinize the absorption imaging method at the microscopic level.\\
\indent To shed more light on this problem we present an oversimplified example of only two atoms located in a spherical harmonic trap. Then, a diagnosis of the system with a light pulse is done through the absorption. Identifying one-photon and a two-photon absorption probabilities as a source of a one-particle density distribution and two-body distribution respectively we study the influence of pulse properties on the results.

\section{Model}

Our method of diagnosing the two-body wave function is based on the absorption of sufficiently well collimated light pulses by atoms - both bosons and fermions. The probabilities of one or two photons being absorbed should be measured for different positions of light beams. Out of the estimated likelihoods we may find one - particle density distribution and two - body distribution. In our studies we are going to ignore spontaneous emission.

The total Hamiltonian of the two contact interacting ultracold particles of equal masses ${m_1} = {m_2}$ absorbing photons from a light beam reads:

\begin{equation}
H = {H_{FS}} + {H_{AF}}.
\end{equation}

The term $H_{FS}$ stands for the free system Hamiltonian. It can be written as:
 
\begin{eqnarray}
{H_{FS}} &=& \underbrace{\sum\limits_{i = 1}^2 {\left( {{T_i}({{\bf{r}}}_i) + {V_t}({{\bf{r}}}_i)} \right)}  + {V_I}({{\bf{r}}}_1 - {{\bf{r}}}_2)}_{H_{S}} \nonumber \\
&\ & + \frac{1}{2}\hbar {\omega _0}\sum\limits_{i = 1}^2{\left( {\left. {\left| e \right.} \right\rangle_i \left\langle {\left. e \right| - } \right.\left. {\left| g \right.} \right\rangle_i \left\langle {\left. g \right|} \right.} \right)},\label{hamfree}
\end{eqnarray}
where ${{T_i}({{\bf{r}}}_i)}=-\frac{\hbar}{2m_i}\nabla^2_{{\bf {r}}_i}$ is the kinetic energy of an atom and ${V_t}({\bf {r}}_i) = \frac{1}{2}m_i\omega {\bf {r}}_i$ a spherical harmonic trapping potential. The short range interaction term for ultracold bosons is expressed by ${V_I}({\bf {r}}) = g\delta_{ps}({\bf {r}})$ with $\delta_{ps}({\bf {r}})$ standing for the pseudo potential which depends on dimensionality, see \cite{Busch1998}. The interactions strength $g$ can be either positive or negative. Its dependence on the scattering length for different dimensions can be found in \cite{Busch1998}.  Note that for two ultra cold fermions in the same spin state ${V_I}({\bf {r}}) =0$. We denote the spatial part of $H_{FS}$ as $H_S$. The last sum in Eq. \eqref{hamfree} describes the internal structure of atoms, which is in a form of a simple two-level model \cite{knight2005}. Here $\left. {\left| g \right.} \right\rangle$ indicates the ground state, $\left. {\left| e \right.} \right\rangle$ stands for the excited state and $\hbar\omega_0$ is the energy difference between two states.\\
\indent The interaction of atoms with a beam is represented by $H_{AF}$ term


\begin{eqnarray}
{H_{AF}} &=& \hbar \lambda \sum\limits_{i = 1}^2 {\left( {\sigma _ + ^i + \sigma _ - ^i} \right) \times}  \nonumber   \\ &\ & \times \left( {\epsilon({{\bf{r}}_i},t){e^{i({{\bf{k}}_L} \cdot {{\bf{r}}_i} - {\omega _L}t)}} + c.c.} \right)  \\
&\approx& \hbar \lambda \sum\limits_{i = 1}^2 {\left( {\sigma _ + ^i{e^{ i({{\bf{k}}_L} \cdot {{\bf{r}}_i} - {\omega _L}t)}} + \sigma _ - ^i{e^{-i({{\bf{k}}_L} \cdot {{\bf{r}}_i} - {\omega _L}t)}}} \right)} \nonumber, \label{hamatfieldfinal}
\end{eqnarray}
where ${\sigma _ \pm }$ are the ladder operators defined as ${\sigma _ + } = \left. {\left| e \right.} \right\rangle \left\langle {\left. g \right|,\quad } \right.{\sigma _ - } = \left. {\left| g \right.} \right\rangle \left\langle {\left. e \right|} \right.$. The expression in the last line of Eq. \eqref{hamatfieldfinal} is obtained by using the Rotating Wave Approximation (RWA) \cite{knight2005}. We assume a weak classical monochromatic beam with an electric field given by ${\bf{E}}({\bf{r}},t) =E_0\left( {\epsilon({\bf{r}},t)e^{i({{\bf{k}}_L} \cdot {\bf{r}} - {\omega _L}t)}} + {\epsilon^{*}({\bf{r}},t)e^{ - i({{\bf{k}}_L} \cdot {\bf{r}} - {\omega _L}t)}}\right)$, where $E_0$ indicates a real-valued magnitude of an amplitude of the electric field, ${\bf k}_L$ a wave vector and $\omega_L$ an angular frequency of the light.  An envelope of a light pulse denoted by $\epsilon({\bf{r}},t)$ is spatially and temporally dependent. We assume a weak intensity of the pulse so that a parameter $\lambda=\frac{dE_0}{\hbar}$ is small as compared to the other terms in the total Hamiltonian. Here $d$ stands for a transition dipole moment of the atom.\\ 
\indent A state vector ${\bf \Psi} ({{\bf r}_1},{{\bf r}_2},t)$ of two atoms within our model can be written in a general form as:
\begin{equation}
{\bf \Psi} ({{\bf r}_1},{{\bf r}_2},t)= \begin{pmatrix} \phi ({{\bf r}_1},{{\bf r}_2},t)\left| {gg} \right\rangle \\  \frac{1}{\sqrt{2}} \left({\chi _1}({{\bf r}_1},{{\bf r}_2},t)\left| {eg} \right\rangle  \pm {\chi _2}({{\bf r}_1},{{\bf r}_2},t)\left| {ge} \right\rangle \right) \\  \eta ({{\bf r}_1},{{\bf r}_2},t)\left| {ee} \right\rangle   \end{pmatrix}
\end{equation}
where $+$ sign in a second line of the state vector corresponds to bosons, whereas $-$ sign to fermions. The time-dependent Shr\"{o}dinger equation  $i\hbar \frac{{\partial {\bf \Psi}}}{{\partial t}} = H{\bf \Psi}$ can be expressed as a system of equations for unknown functions  $ \phi$, $ \chi_1$, $ \chi_2 $ and $ \eta$ by:
\begin{equation}
\left\{
\begin{aligned}
   i\hbar {\mathop \phi \limits^.} &= \left( {{H_S} - \hbar {\omega _0}} \right) \phi   +  \frac{\hbar \lambda}{{\sqrt 2 }}  \left( {\epsilon^*}({{\bf{r}}_1},t){e^{ - i({{\bf{k}}_{\bf{L}}} \cdot {{\bf{r}}_{\bf{1}}} - {\omega _L}t)}}{\chi _1} \right. \\  
  &  \qquad \qquad \qquad \qquad \pm \left. {\epsilon^*}({{\bf{r}}_2},t){e^{ - i({{\bf{k}}_{\bf{L}}} \cdot {{\bf{r}}_2} - {\omega _L}t)}}{\chi _2} \right) \\
    i\hbar {\mathop \chi \limits^.}_1 &= {H_S}{\chi _1} + \sqrt 2 \hbar \lambda \left( \epsilon({{\bf{r}}_1},t){e^{i({{\bf{k}}_{\bf{L}}} \cdot {{\bf{r}}_1} - {\omega _L}t)}}\phi \right. \\ & \qquad \qquad \qquad \qquad  \left.  +   {\epsilon^*}({{\bf{r}}_2},t){e^{ - i({{\bf{k}}_{\bf{L}}} \cdot {{\bf{r}}_2} - {\omega _L}t)}}\eta  \right)  \\
  i\hbar {\mathop \chi \limits^.}_2 &= {H_S}{\chi _2} \pm \sqrt 2 \hbar \lambda \left( \epsilon({{\bf{r}}_2},t){e^{i({{\bf{k}}_{\bf{L}}} \cdot {{\bf{r}}_2} - {\omega _L}t)}}\phi \right. \\ & \qquad \qquad \qquad \qquad  \left.  +   {\epsilon^*}({{\bf{r}}_1},t){e^{ - i({{\bf{k}}_{\bf{L}}} \cdot {{\bf{r}}_1} - {\omega _L}t)}}\eta  \right) \\
 i\hbar \mathop \eta \limits^.  &= \left( {{H_S} + \hbar {\omega _0}} \right)\eta  + \frac{\hbar \lambda }{{\sqrt 2 }}\left( \epsilon({{\bf{r}}_2},t){e^{i({{\bf{k}}_{\bf{L}}} \cdot {{\bf{r}}_2} - {\omega _L}t)}}{\chi _1} \right. \\ &  \qquad \qquad \qquad \qquad  \left. \pm \epsilon({{\bf{r}}_1},t){e^{i({{\bf{k}}_{\bf{L}}} \cdot {{\bf{r}}_1} - {\omega _L}t)}}{\chi _2} \right)
 \end{aligned}\label{eqs}
\right.
\end{equation}

The above system of equations may be solved approximately in the following way. As it was mentioned before we consider a very weak driving to neglect a depletion of the initial state. We also assume that initially two atoms are in the internal ground states, namely that for $t=0$ 
a state vector ${\bf \Psi} ({{\bf r}_1},{{\bf r}_2},0)=\phi ({{\bf r}_1},{{\bf r}_2},0)\left| {gg} \right\rangle$. Therefore we assume that during the interaction between the system and the light the state vector remains almost unchanged, that is to say $\left| \phi  \right| \gg \left| {{\chi _1}} \right|,\left| {{\chi _2}} \right| \gg \left| \eta  \right|$ for the duration of the pulse. Then, introducing the interaction picture by following substitutions  $\phi  \to \quad {e^{i{\omega _0}t}}\phi$, $\eta  \to \quad  {e^{ - i{H_S}t/\hbar }}{e^{ - i{\omega _0}t}}\eta$ and ${\chi _{1(2)}} \to \quad {e^{ - i{H_S}t/\hbar }}{\chi _{1(2)}}$ we obtain the final equations
\begin{equation}
\left\{
\begin{aligned}
    \phi &=  {e^{ - i{H_S}t/\hbar }}  \phi({\bf {r}}_1,{\bf {r}}_2,0) \\
     {\mathop \chi \limits^.}_1&= -i\sqrt 2 \lambda {e^{  i{H_S}t/\hbar }} \epsilon({{\bf{r}}_1},t){e^{i{{\bf{k}}_{\bf{L}}} \cdot {{\bf{r}}_1}}}{e^{ - i\Delta t}} \phi \\
  {\mathop \chi \limits^.}_2 &= \mp i \sqrt 2 \lambda {e^{  i{H_S}t/\hbar }} \epsilon({{\bf{r}}_2},t){e^{i{{\bf{k}}_{\bf{L}}} \cdot {{\bf{r}}_2}}} {e^{ - i\Delta t}}\phi \\
\mathop \eta \limits^.  &=  -i \frac{1}{{\sqrt 2 }} \lambda \left( {e^{  i{H_S}t/\hbar }} \epsilon({{\bf{r}}_2},t){e^{i{{\bf{k}}_{\bf{L}}} \cdot {{\bf{r}}_2}}}{e^{ - i\Delta t}}{e^{ - i{H_S}t/\hbar }} {\chi _1} \right. \\ &  \left. \qquad \pm{e^{  i{H_S}t/\hbar }} \epsilon({{\bf{r}}_1},t){e^{i{{\bf{k}}_{\bf{L}}} \cdot {{\bf{r}}_1}}}{e^{ - i\Delta t}}{e^{ - i{H_S}t/\hbar }} {\chi _2} \right)  
 \end{aligned}
\right.\label{eqsfinal}
\end{equation}
where we define a detuning by  $\Delta=\omega_L-\omega_0$. As we may note the final form of the above system of equations comes directly from the fact that in a general case $\left[ {\epsilon({\bf {r}}_i){e^{i{{\bf{k}}_{\bf{L}}} \cdot {{\bf{r}}_i}}},{H_S}} \right] \ne 0$.
\section{Solutions}
The analytical solutions of the spatial Hamiltonian $H_{S}$ are well known both for two ideal bosons or fermions (${V_I}({\bf {r}}) = 0$) and for two interacting ultra cold bosons \cite{Busch1998}. We assume a rectangle pulse envelope. When the light is on $ \epsilon({{\bf{r}}},t)=\epsilon({{\bf{r}}})$. Thus it is possible to solve Eq. \eqref{eqsfinal} analytically. Without loss of generality we choose the initial state as an eigenvector of $H_S$, namely that $\phi ({{\bf r}_1},{{\bf r}_2},0)=\phi_n({{\bf r}_1},{{\bf r}_2})$ with an index $n$ indicating the nth eigenvector in a chosen basis. Then using Dirac notation and a formula ${e^{ - i{H_S}t/\hbar }} = \sum\limits_i {{e^{ - i{E_i}t/\hbar }}\left| {{\phi _i}} \right\rangle \left\langle {{\phi _i}} \right|} $ with $E_i$ standing for the i-th eigenvalue we find a general solution for $ \chi_1$, $ \chi_2 $ and $ \eta$ as

\begin{equation}
\left\{
\begin{split}
   \phi &=  {e^{ - i{E_n}t/\hbar }}\left| {{\phi _n}} \right\rangle\\
     \chi_1&= -\sqrt 2 \lambda \sum\limits_i {\epsilon_{in}\frac{\left({e^{  i{\tilde{\Delta}_{in}}t}}-1\right)}{\tilde{\Delta}_{in}}  \left| {{\phi _i}} \right\rangle}  \\
     \chi_2&= \mp\sqrt 2 \lambda \sum\limits_i {\tilde{\epsilon}_{in}\frac{\left({e^{  i{\tilde{\Delta}_{in}}t}}-1\right)}{\tilde{\Delta}_{in}}  \left| {{\phi _i}} \right\rangle}  \\
\eta  &=  -\lambda^2 \sum\limits_{i,k} \frac{\tilde{\epsilon}_{ik}{\epsilon}_{kn}+{\epsilon}_{ik}\tilde{\epsilon}_{kn}}{\tilde{\Delta}_{kn}}\times \\ & \quad \left(\frac{{e^{  i{\tilde{\Delta}_{ik}}t}}-1}{\tilde{\Delta}_{ik}}-\frac{{e^{  i\left({\tilde{\Delta}_{ik}}+{\tilde{\Delta}_{kn}}\right)t}}-1}{\tilde{\Delta}_{ik}+\tilde{\Delta}_{kn}}\right)  \left| {{\phi _i}} \right\rangle
 \end{split}
\right.\label{solution}
\end{equation}
where $\epsilon_{ij}=\left \langle {{\phi _i}} \right|\epsilon({\bf{r}}_1){e^{i{{\bf{k}}_{\bf{L}}} \cdot {{\bf{r}}_1}}}\left| {{\phi _j}} \right\rangle$, $\tilde{\epsilon}_{ij}=\left \langle {{\phi _i}} \right|\epsilon({\bf{r}}_2){e^{i{{\bf{k}}_{\bf{L}}} \cdot {{\bf{r}}_2}}}\left| {{\phi _j}} \right\rangle$ and obviously $\epsilon_{ij}=\tilde{\epsilon}_{ij}$. Here we define a scalar product by $ \langle {{\phi _i}} \left| {{\phi _j}} \right\rangle=\int{d{\bf{r}}_1d{\bf{r}}_2{\phi^*_i}({\bf{r}}_1,{\bf{r}}_2){\phi_j}({\bf{r}}_1,{\bf{r}}_2)}$. The generalized energy difference between i-th and j-th states reads $\tilde{\Delta}_{ij}=\Delta_{ij}-\Delta=\frac{E_i-E_j}{\hbar}-\Delta$. Note that for resonant terms i.e. $\tilde{\Delta}_{ij}=0$ in any sum of Eq. \eqref{solution} its proper element has to be evaluated by taking a limit $\tilde{\Delta}_{ij}\rightarrow 0$. A resonant term behaves as $t$ for $\chi_1$, $\chi_2$ and as $t^2$ for $\eta$. For the clarity of our argumentation hereafter we will take a resonant case with $\Delta=0$, but our conclusions will hold also for $\Delta \neq 0$.\\
\indent The probabilities of having one or two photons absorbed by atoms are easily defined by $P_1(t)={\left| {\chi _1}\left| {eg} \right\rangle  \pm {\chi _2} \left| {ge} \right\rangle \right|}^2={\left| {\chi _1} \right|}^2+{\left| {\chi _2} \right|}^2 $ and $P_2(t)={\left| {\eta}\left| {ee} \right\rangle \right|}^2={\left| {\eta} \right|}^2$. After some straightforward calculations they can be expressed as:

\begin{align}
P_1(t) &=2{\lambda ^2}{\left| {{\epsilon _{nn}}} \right|^2}{t^2} + 4{\lambda ^2}\sum\limits_{i \ne n} {\frac{{{{\left| {{\epsilon _{in}}} \right|}^2}}}{{{\Delta ^2}_{in}}}\left( {1 - \cos ({\Delta _{in}}t)} \right)}\label{p1} \\
P_2(t) &={\lambda ^4}{\left| {{\epsilon _{nn}}} \right|^4}{t^4}  + 4\lambda^4\sum\limits_{\scriptstyle i \ne n \hfill \atop 
  \scriptstyle k,k' \hfill}  {\epsilon^*_{ik'}\epsilon^*_{k'n}\epsilon_{ik}\epsilon_{kn}}p_{ikk'n}(t)\label{p2},
\end{align}

where $p_{ikk'n}(t)$ is of order $o(t^4)$. Its definition can be find in the footnote \footnote{The full expression for $p_{ikk'n}(t)$ is given by:
\begin{equation}\label{eqapp}
\begin{split}
p_{ikk'n}(t)=&\frac{1}{\Delta_{k'n}\Delta_{kn}}\times \\
&\left(\frac{{e^{  i{\Delta_{ik}}t}}-1}{{\Delta}_{ik}}-\frac{{e^{  i\left({{\Delta}_{ik}}+{{\Delta}_{kn}}\right)t}}-1}{{\Delta}_{ik}+{\Delta}_{kn}}\right) \times \\  & \left(\frac{{e^{  -i{\Delta_{ik'}}t}}-1}{{\Delta}_{ik'}}-\frac{{e^{  -i\left({{\Delta}_{ik'}}+{{\Delta}_{k'n}}\right)t}}-1}{{\Delta}_{ik'}+{\Delta}_{k'n}}\right) 
\end{split}
\end{equation}
Note that for resonant terms i.e. ${\Delta}_{ij}=0$ in any sum of Eq. \eqref{eqapp} its proper element has to be evaluated by taking a limit ${\Delta}_{ij}\rightarrow 0$. By using Eq. \eqref{eqapp} it is easy to check that the sum in Eq. \eqref{p2} is a real function of time.}. A short time characteristic of the probabilities, when $t\ll \Delta_{in}^{-1}$ with $i$ corresponding to the nearest eigenvalue to $n$, reads:

\begin{align}
P_1(t) &\approx 2{\lambda ^2}\left \langle {{\phi _n}} \right|\left|\epsilon({\bf{r}}_1)\right|^2\left| {{\phi _n}} \right\rangle{t^2}, \quad t\rightarrow 0 \label{p1s} \\
P_2(t) &\approx{\lambda ^4}\left \langle {{\phi _n}} \right|\left|\epsilon({\bf{r}}_1)\right|^2\left|\epsilon({\bf{r}}_2)\right|^2\left| {{\phi _n}} \right\rangle{t^4}, \quad t\rightarrow 0 \label{p2s}.
\end{align}
 The analysis of Eq. \eqref{p1}, \eqref{p2}, \eqref{p1s} and \eqref{p2s} reveals an intriguing discrepancy between the short time and the long time behaviour of the probabilities. First of all, the long time probabilities depend on couplings between an actual state of the system and different eigenstates that occur because, for the experimental relevance, beam width must be narrower than a characteristic system width. This fact automatically leads to the conclusion that for longer pulses the information about the actual state of the system is blurred. Secondly, although the dominant terms are of the same order in both situations, the coefficients determining their magnitude are not. For the short time the probabilities coefficients are related to the intensity $\left|\epsilon({\bf{r}})\right|^2 $. Note that for $P_1(t), \quad t \rightarrow 0$ the coefficient in front of $t^2$ can be rewritten as $\int{d{\bf{r}}{\rho}({\bf{r}})\left|\epsilon({\bf{r}})\right|^2}$ with a one-particle density ${\rho}({\bf{r}})=\int{d{\bf{r}}'\left|\phi_n({\bf{r}},{\bf{r}}') \right|^2 }$ which is a very intuitive result. On the other hand the coefficients for the long time depends on the amplitude of the pulse proportional to $\epsilon({\bf{r}}){e^{i{{\bf{k}}_{\bf{L}}} \cdot {{\bf{r}}}}}$ rather than to its intensity alone. In the next section we are going to show the most striking examples of the above differences.
\section{Results}
In this section we present results obtained within our model which are mimicking an experiment diagnosing a quantum state of two ultra cold atoms. We restrict our findings to a quasi-1D system which captures all essentials features of our model and provides with a clear picture. In a real experiment it corresponds to cigar-shaped traps with a very strong transverse confinement. We send a probing light pulses along transverse direction z which is related to an electric field ${\bf{E}}({\bf{r}},t) =E_0\left( {\epsilon(x)e^{i(k_L z - {\omega _L}t)}} + {\epsilon^{*}(x)e^{ - i(k_L z - {\omega _L}t)}}\right)$ with $k_L=\left| {\bf{k}}_L \right|$. We may also assume that $1 /k_L$ is much bigger than a typical transverse length of a probe so that a driving term $e^{\pm i(k_L z )}$ may be neglected. As the initial state $\phi(x_1,x_2,0)$ we select the ground state both for ideal fermions or bosons and interacting bosons.
\subsection{One-particle density function}
\begin{figure}
\centering
\includegraphics[width=0.49\textwidth]{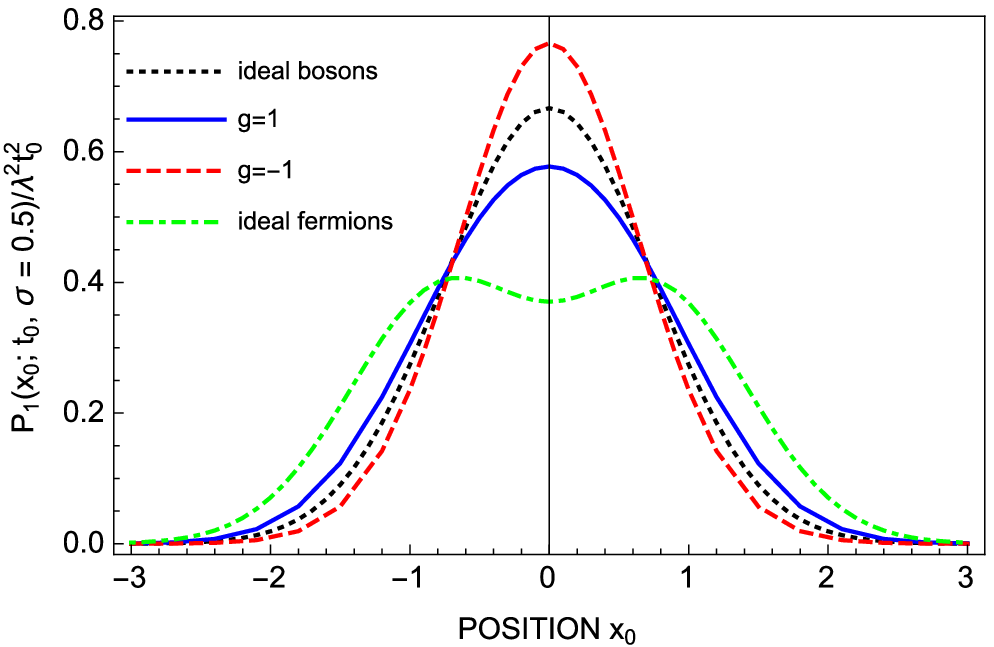}
\caption{\label{fig:fro}}One-photon absorption probability $P_1(x_0; t_0, \sigma)/\lambda^2t_0^2$ as a function of a beam center $x_0$.  A time $t_0$ was chosen so that Eq. \eqref{p1} and Eq. \eqref{p1s} agree with each other i.e. $t_0=0.0001\Delta_{10}^{-1}$, where $\Delta_{10}$ is related to the energy difference between the ground and the first excited state. The straight blue line corresponds to two ideal bosons, the straight red line comes from two interacting repulsively bosons, the black dashed line is related to two interacting attractively bosons and the green dashdotted line to two fermions. The oscillatory units are used.
\end{figure}

\begin{figure}
\centering
\includegraphics[width=0.49\textwidth]{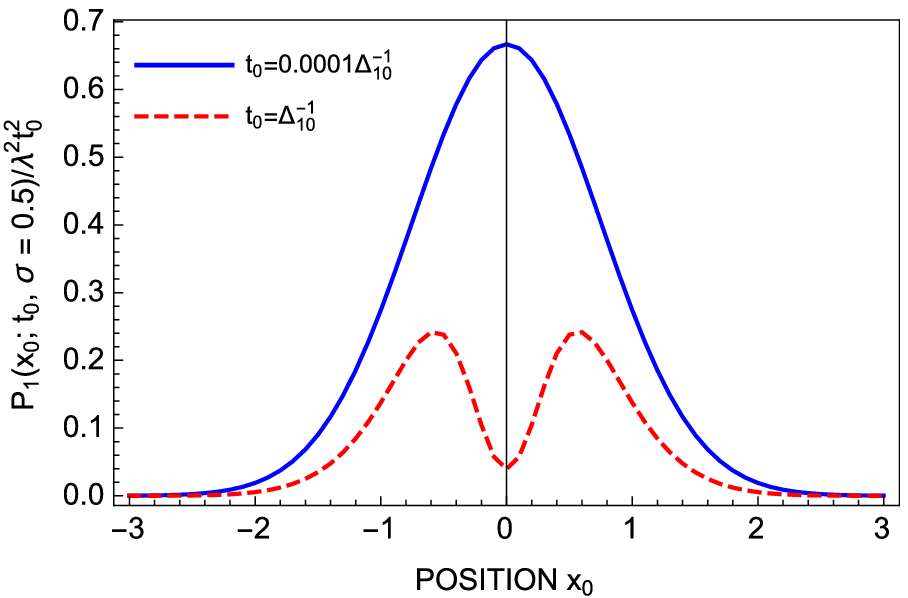}
\caption{\label{fig:fcompro}}One-photon absorption probability $P_1(x_0; t_0, \sigma)/\lambda^2t_0^2$ as a function of a beam center $x_0$ for two ideal bosons. The straight blue line corresponds to time $t_0=0.0001\Delta_{10}^{-1}$,  whereas the red dashed to time $t_0=\Delta_{10}^{-1}$. The oscillatory units are used.
\end{figure}

In order to find a one-particle density function we use a single pulse with $\epsilon (x;{x_0},\sigma ) = \frac{1}{{\sigma \sqrt \pi  }}{e^{ - {{\left( {x - {x_0}} \right)}^2}/{\sigma ^2}}}$. Then using a short-time characteristic of $P_1(t)$ expressed by Eq. \eqref{p1s} we evaluate a probability of one-photon absorption as a function of a position of the pulse center $x_0$. We compile our results for two ideal or interacting bosons and two fermions in Fig. \ref{fig:fro}. As we may note by comparing with the well known analytical solutions of $H_S$ for the ground state the one-photon absorption diagnosis gives a direct access to the one-particle density distribution $\rho(x)$ defined in the preceding section. A clear difference between interacting and non-interacting case is seen as well as between bosons and fermions. A repulsive system distribution is wider than that of an ideal gas, while an attractive system is narrower than the ideal one.\\
\indent The analysis of Eq. \eqref{p1} shows that using a pulse that is too long may affect a measured one-particle density profile. For pulses with complicated wave fronts the coefficients in the Eq. \eqref{p1} would differ significantly from these of the Eq. \eqref{p1s}. To illustrate the unwanted field amplitude dependence of the result for long pulses we choose an extreme example of a pulse with cross-section given by $\epsilon (x;{x_0},\sigma ) = \frac{1}{{\sigma \sqrt \pi  }}{e^{ - {{\left( {x - {x_0}} \right)}^2}/{\sigma ^2}}}{\rm sgn(x)}$ with ${\rm sgn(x)}$ staying for the sign function. A comparison between the one-particle density profiles for the ideal bosons for two different pulse durations $t_0=0.0001\Delta_{10}^{-1}$ and $t_0=\Delta_{10}^{-1}$, where $\Delta_{10}$ is related to the energy difference between the ground and the first excited state, is presented in Fig. \ref{fig:fcompro}. For the result based on Eq. \eqref{p1} we truncate the sum at $i=20$ ensuring that adding another eigenstate would not change the result up to 1\% accuracy. A striking difference can be observed. The density profile obtained after a measurement with a long pulse has nothing in common with the actual one-particle density. It is a clear indication that a diagnosis of a few-body quantum state can be highly biased for longer pulses.

\subsection{Two-body wave function}
\begin{figure*}
\centering
\includegraphics[width=0.75\textwidth]{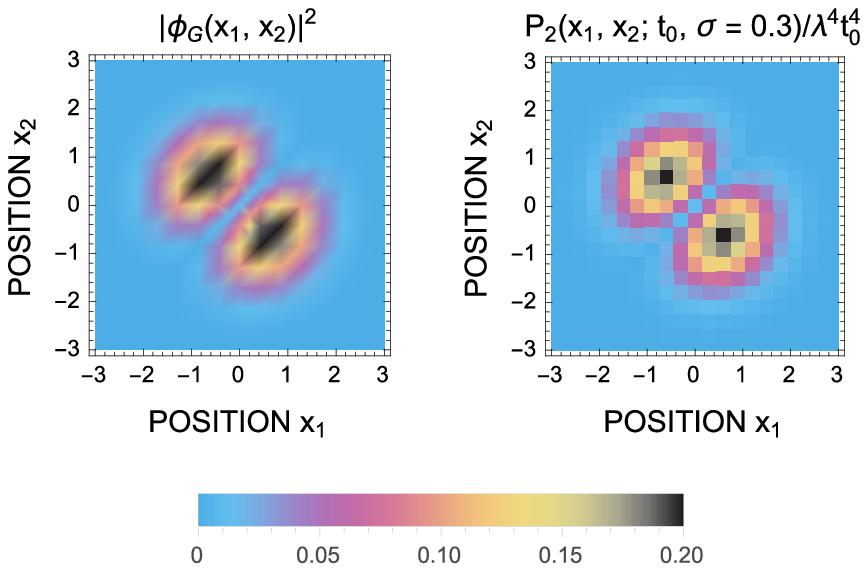}
\caption{\label{fig:wavfun}} Two-photon absorption probability $P_2(x_1, x_2; t_0, \sigma)/\lambda^4t_0^4$ for two interacting bosons with $g=6$  as a function of a beam positions $x_1$ and $x_2$ for $t_0=0.0001\Delta_{10}^{-1}$ after subtracting a single beam two-photon absorption (right). The oscillatory units are used. Left plot shows the analytical solution of $\left| \phi_G(x_1, x_2)\right|^2$ for $g=6$.
\end{figure*}

Measuring a two-photon absorption probability a two-body wave function can be diagnosed. To achieve that one has to use a pulse with a double-focused envelope, namely with $\epsilon (x;{x_1},{x_2},\sigma ) = \frac{1}{{\sigma \sqrt \pi  }}\left( {{e^{{{\left( {x - {x_1}} \right)}^2}/{\sigma ^2}}} + {e^{{{\left( {x - {x_2}} \right)}^2}/{\sigma ^2}}}} \right)$. Then a coefficient in Eq. \eqref{p2s} reads 
\begin{equation*}
\begin{split}
& \left \langle {{\phi _G}} \right|\left|\epsilon(x; x_1, x_2, \sigma)\right|^2\left|\epsilon(y; x_1, x_2, \sigma)\right|^2\left| {{\phi _G}} \right\rangle \approx \\
& \qquad \left \langle {{\phi _G}} | 2\left|\epsilon(x; x_1, \sigma)\right|^2\left|\epsilon(y;  x_2, \sigma)\right|^2 +  \left|\epsilon(x; x_1, \sigma)\right|^2 \right. \times \\
&\qquad \qquad \quad \left. \left|\epsilon(y;  x_1, \sigma)\right|^2  +\left|\epsilon(x; x_2, \sigma)\right|^2\left|\epsilon(y;  x_2, \sigma)\right|^2 | {{\phi _G}} \right\rangle 
\end{split}
\end{equation*}
with $\left| {{\phi _G}} \right\rangle$ denoting the ground state of $H_{S}$. Here we neglect the interference terms like ${e^{{{\left( {x - {x_1}} \right)}^2}/{\sigma ^2}}}{e^{{{\left( {x - {x_2}} \right)}^2}/{\sigma ^2}}}$ that in experiment can be realized either by ensuring $\left| x_1-x_2\right| > 3\sigma$ or by introducing a phase difference between two pulses and averaging over many measurements. The last two terms of a sum in the above equation are corresponding to processes where two photons were absorbed at the same space point. The probability of such a process should be measured independently and then subtracted from the total result of the two-photon absorption. The easiest way to notice that is by assumption that $\left|\epsilon(x; x_1, \sigma)\right|^2\approx \delta (x-x_1)$ which makes 
$\left \langle {{\phi _G}} \right|\left|\epsilon(x; x_1, x_2, \sigma)\right|^2\left|\epsilon(y; x_1, x_2, \sigma)\right|^2\left| {{\phi _G}} \right\rangle \approx 2\left|\phi_G( x_1, x_2)\right|^2+\left|\phi_G( x_1, x_1)\right|^2+\left|\phi_G( x_2, x_2)\right|^2$.\\
\indent We sum up our considerations with an example of two repulsive bosons with interaction strength $g=6.$. The total two-photon absorption probability $P_2(t)$ was found for $t_0=0.0001\Delta_{10}^{-1}$ and for several beam positions $x_1$ and $x_2$. Then we subtract from it the probabilities of a single pulse two-photon absorption. Finally we compare our findings with the analytical solution for $\phi_G(x,y)$ which is plotted in Fig. \ref{fig:wavfun}. It is a straightforward observation that we reconstructed the actual two-body wave function density within our model.

\begin{figure*}
\centering
\includegraphics[width=0.75\textwidth]{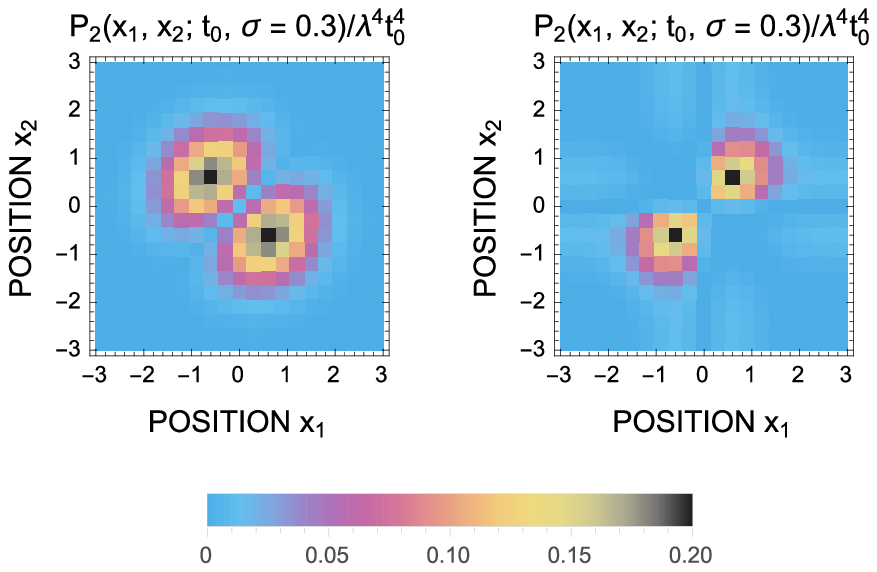}
\caption{\label{fig:bias}}Two-photon absorption probability $P_2(x_1, x_2; t_0, \sigma)/\lambda^4t_0^4$ for two interacting bosons with $g=6$ as a function of a beam positions $x_1$ and $x_2$ for $t_0=0.0001\Delta_{10}^{-1}$ (left) and $t_0=\Delta_{10}^{-1}$ (right) after subtracting a single beam two-photon absorption. The oscillatory units are used.
\end{figure*}

Analogously to the previous subsection the results for a long pulse in a time domain when Eq. \eqref{p2} holds may lead to a wrong two-body wave function. One more time we use the example of the highly modified wave front with 
 $\epsilon (x;{x_1},{x_2},\sigma )=\frac{1}{{\sigma \sqrt \pi  }}\left( {{e^{{{\left( {x - {x_1}} \right)}^2}/{\sigma ^2}}} + {e^{{{\left( {x - {x_2}} \right)}^2}/{\sigma ^2}}}} \right) {\rm sgn(x)}$. The resulting two-photon probability absorption $P_2(t)/\lambda^4t^4$ for $t_0=0.0001\Delta_{10}^{-1}$ and $t_0=\Delta_{10}^{-1}$ for two interacting bosons with $g=6.$ as a function of beam positions can be found in Fig. \ref{fig:bias}. For the result based on Eq. \eqref{p2} we truncate the sum at $i=20$ ensuring that adding another eigenstate would not change the result up to 1\%.  accuracy. Our findings stress the fact that a pulse duration in an experimental diagnosis should be chosen very carefully.

\section{Conclusions}

In conclusions, we studied a simple model of diagnosing a two-body state with light for interacting or ideal bosons and fermions. We demonstrated that results of an experiment based on our theory would crucially depend on a pulse duration. For a sufficiently short pulses we correctly estimate with our measurement the actual one-particle density function and the two-body wave function. For longer pulses a hypothetical experimental findings would be highly biased. The main reason for that is that the calculated probabilities of one-photon or two-photons absorptions are related to the intensity of the light beam for sufficiently short time, whereas for longer time they depend on the amplitude of a pulse. The structure of Eq. \eqref{p1} and Eq. \eqref{p2} can be understood that the probability of the absorption in a space point is strongly blurred by free evolution of the initial state. Our results can be generalized to systems containing more particles which allows to investigate an experimental procedure of diagnosing higher-order correlation functions from the many-body perspective.
\acknowledgements
RO and KRz acknowledge the support of (Polish) National Science Center grant no. 2015/19/B/ST2/02820
\bibliographystyle{eplbib.bst}
\bibliography{articlebib}
\end{document}